\documentclass[12pt]{amsart}
\usepackage{amssymb,latexsym,graphics}
\usepackage{times}

\usepackage{times}

\title[Yeast Protein Interactome]{Yeast Protein Interactome topology\\ provides framework for\\ coordinated-functionality} 

\author
[ ] {Andr\'{e} X. C. N. Valente$^{1*}$, Michael E. Cusick$^{2}$
\\
$^{1}$ 
Biometry Research Group,\\ National Cancer Institute,\\ National Institutes of Health,\\ Dept. of Health and 
Human Services\\
Bethesda, MD 20892 USA\\
 $^{2}$
Center for Cancer Systems Biology and Dept. of Cancer Biology\\ Dana-Farber Cancer Institute\\ and Dept. of Genetics, Harvard Medical School\\ Boston, MA 02115 USA\\
$^\ast$To whom correspondence should be addressed; E-mail: andre@deas.harvard.edu} 

\begin{document} 
\maketitle


{\bf Summary}
\vspace{0.1cm}

{\bf The architecture of the network of protein-protein physical interactions in {\em Saccharomyces cerevisiae} is exposed through the combination of two complementary theoretical network measures, betweenness centrality and `Q-modularity'.
The yeast interactome is characterized by well-defined topological modules connected via a small number of inter-module protein interactions.
Should such topological inter-module connections turn out to constitute a form of functional coordination between the modules, we speculate that this coordination is occurring typically in a pair-wise fashion, rather than by way of high-degree hub proteins responsible for coordinating multiple modules. The unique non-hub-centric hierarchical organization of the interactome is not reproduced by gene duplication-and-divergence stochastic growth models that disregard global selective pressures.
}

\vspace{0.5cm}
{\bf Introduction}

\begin{figure} 
\scalebox{0.5}{\includegraphics{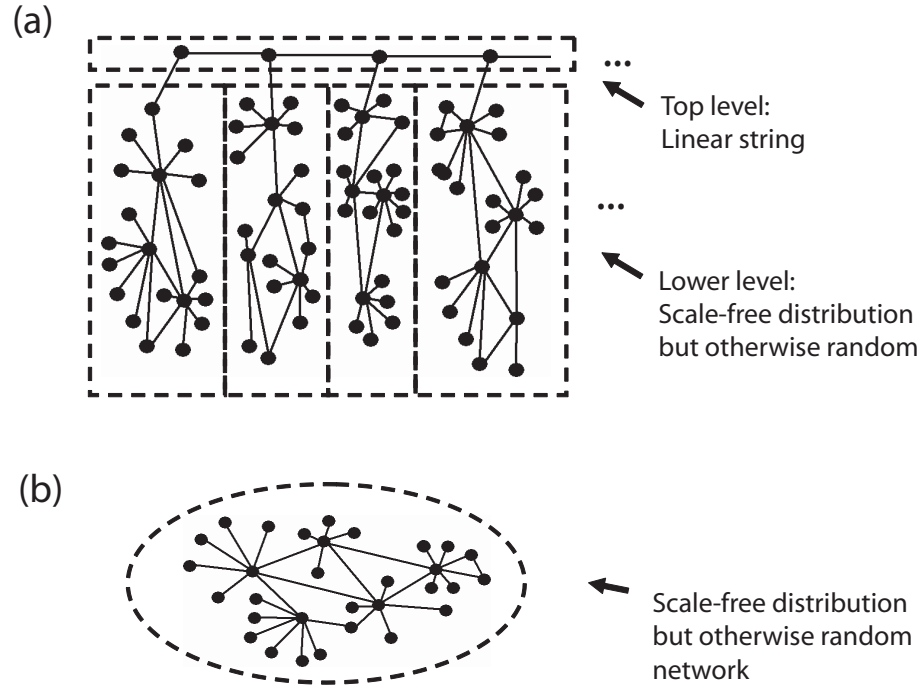}} 
\caption{{\em Hidden intrinsic hierarchy in networks.} {In this example sketch, network (a) has two hierarchical levels: it consists of a linear string of nodes at the top level that connect to a lower level set of subgraphs possessing scale-free degree distributions, but that are otherwise random. Network (b) has only one hierarchical level, also with a scale-free degree distribution and random in other respects. Arguably, for many applications the difference between topology (a) and topology (b) is of relevance. Yet, an analysis based on common measures such as degree distribution~\cite{BarabasiBioReview,NewmanReview}, clustering coefficient as a function of degree~\cite{BarabasiBioReview,BarabasiHierarchy}, or degree correlation measures~\cite{Sneppen}, to name a few, would indicate networks (a) and (b) to be topologically identical. This is due to the much larger number of nodes at the lower hierarchical level statistically overwhelming, and therefore hiding, the top level structure.}}
\label{HierarchyFig}
\end{figure}

The set of all physical protein-protein interactions in a cell -- the interactome -- presents a foundational picture for Biology, sitting at the lowest level of description at which it is possible to have an holistic view of a cell rather then just an isolated study of its individual components. 
In this article, we make a small contribution to the ongoing effort to understand the global architecture of this fundamental physical network~\cite{MarcotteInsight, InteractomeVisualization,VidalAtlas,  BooneReview,TyersReview,UetzReview,AlbertsMachines,YeastToHuman,SupertiFugaReview,Jackie}.
The interactome can be represented in an abstract way as a network of nodes connected by links, where nodes stand for proteins and links for direct physical interactions between proteins. In recent years, there has been much interest in applying statistical mechanics to the study of such complex networks~\cite{BarabasiBioReview, NewmanReview}. 
However, the validity of such an approach is always conditional on the fundamental assumptions of statistical mechanics being satisfied.
For instance, many statistical measures will not be able to distinguish between a network with an intrinsic hierarchical topology and one without it (Figure~\ref{HierarchyFig}).

\vspace{0.4cm}

{\bf Results and Discussion}

Our approach sidesteps the limitations alluded to in the preceding paragraph and focuses directly on how the  interactome topology relates to two broad biological concepts. The first of these, hierarchical organization, is in this context the notion that there may exist a hierarchy in the role of proteins~\cite{Jackie}. On one hand, there are proteins that perform very specific, local functions, relevant only within the context of a particular biological process. On the other hand, some proteins may possess a global, high level role, perhaps acting as mediators of distinct biological processes. To study the topological hierarchy in the interactome, we use the graph theoretical betweenness centrality measure~\cite{BetweennessOriginal,SuiHuang,Kern} (Supp. Mat.). Betweenness centrality (denoted `traffic', henceforth) for a node is the total number of shortest paths (between any two other nodes) in the network that pass through that node. A high traffic value for a protein therefore correlates with that protein being topologically central in the interactome.
The second of these is the concept of biological functional modularity~\cite{MurrayModular}. In the context of proteins, at the extreme, this takes the form of protein machines performing specific functions in a cell~\cite{AlbertsMachines}. More generally, it consists of an expectation that the density of protein-protein interactions will rise as we zoom into an increasingly functionally related set of proteins. To assess modularity in the interactome topology, we use the `Q-modularity' measure of Newman~\cite{NewmanFast}, which assigns a modularity score, Q, to any given partition of the network into modules. The modularity Q is defined as the difference between the ratio $(\mbox{intra-module edges})/(\mbox{total edges})$ for the network in question, and the expected value for this ratio if edges in the network were randomized, subject to every node maintaining its original degree.
 We use the algorithm of Clauset et al.~\cite{NewmanLargeNets} to find an interactome partition into modules that corresponds to a large Q value.

We now explain how to produce an {\em interactome polar map} (Figure 2) by combining the information contained in the modularity and traffic analyses.
The position of every protein in the map is specified in terms of its radial and angular coordinates. The radial coordinate is a function of its traffic~\cite{BrandesCentrality}. More precisely, it is proportional to $\log \left(\mbox{max traffic}/\mbox{protein's traffic}\right)$, where `max traffic' is the maximum node traffic in the network (Supp. Mat.).
A logarithmic scale is used due to the long tail of the traffic distribution~\cite{KimBetweenness}.
The protein angular coordinates are assigned such that all proteins in the same module fall within the same angular range (Supp. Mat.). This way, an interactome map is created where topologically increasingly central proteins are radially increasingly closer to the center of the map, while angular sectors correspond to topological modules. To determine the circular ordering of the modules in the map, we introduce a {\em Ring Ordering Algorithm} that, based on the interactome inter-module connectivity, attempts to place closer to each other, to the extent that it is possible, modules that are more topologically related (Supp. Mat.).

\begin{figure}
\scalebox{0.95}{\includegraphics{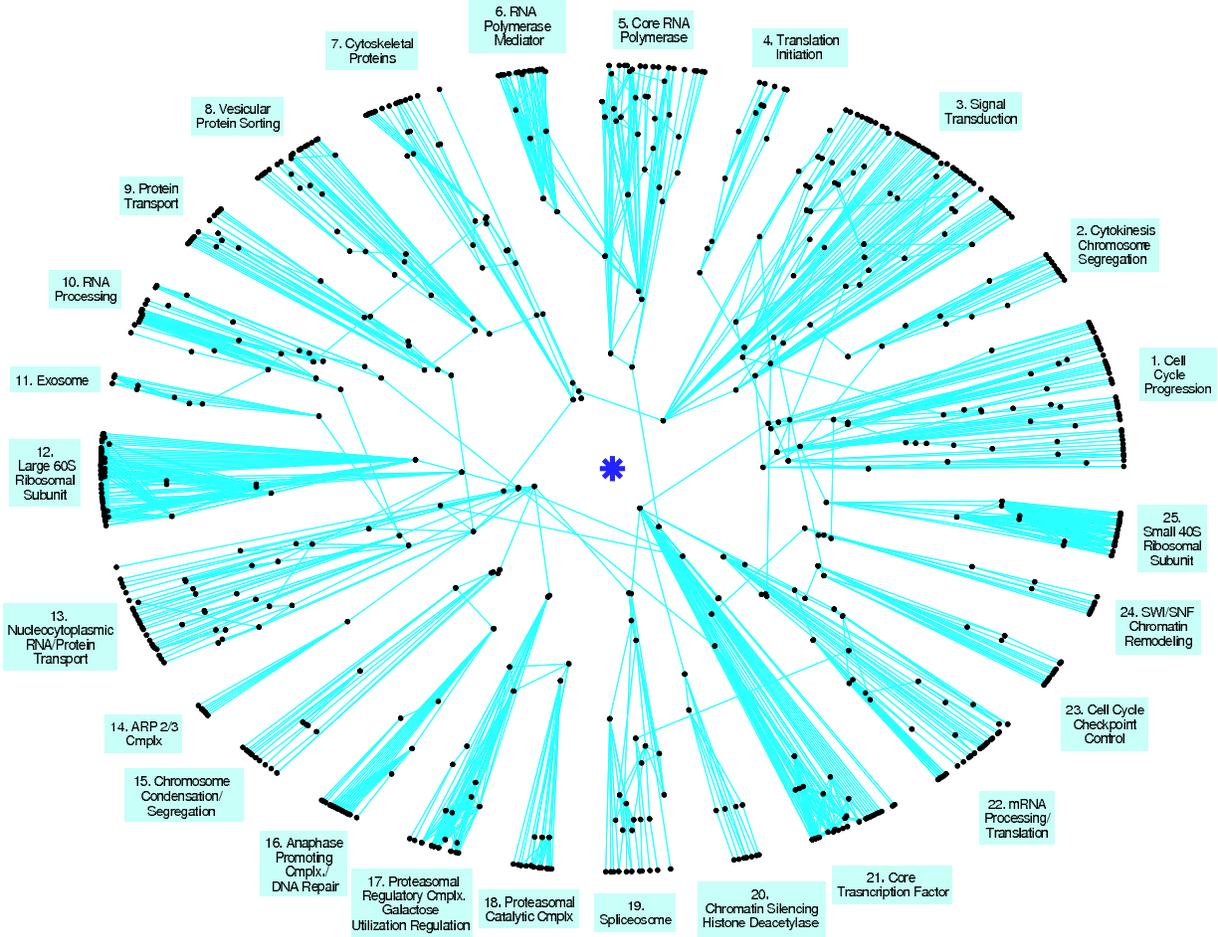}} 
\caption{{\em Saccharomyces cerevisiae interactome polar map.} {The map is constructed, in an unsupervised manner, based solely on protein-protein interaction data. The module captions (blue text boxes) were manually chosen, a posteriori, to reflect the biological role of each module. The map suggests a `coordinated-functionality' architecture for the interactome, arguably an ideal framework for the cell to physically implement the concept of distinct, yet coordinated, biological functional modules. This would be a {\em pair-wise-coordination}, as inter-module physical interactions occur in a pair-wise fashion: of the 76 proteins that possess inter-modular connections, only 4 connect their module to more than a single other module (TAF25 in module \#21 and SRP1 in module \#13 have links to four other modules, while NUP1 in module \#13 and CLB2 in module \#1 have links to two other modules). 
 The map is based on the higher confidence FYI protein-protein interaction data set~\cite{Jackie}, consisting of interactions validated either through small-scale experiments or through at least two distinct procedures. The giant connected component, shown here, consists of 741 proteins and 1752 protein-protein interactions.}}
\label{MainFig}
\end{figure}

We apply this analysis and discuss its implications in the context of the {\em Saccharomyces cerevisiae} interactome. 
The interactome data set we use is the higher confidence `filtered yeast interactome' (FYI)~\cite{Jackie}, consisting of interactions supported either by small-scale screens as reported in the MIPS database~\cite{MIPSdatabase} or by at least two distinct methods from amongst i) high-throughput yeast two-hybrid experiments~\cite{UetzY2H,ItoY2H}, ii) computational predictions based on gene co-occurrence~\cite{InSilico,InSilico2}, gene neighborhood~\cite{InSilico,InSilico3} or gene fusion~\cite{InSilico,InSilico2}, iii) high-throughput affinity purification/mass spectrometric protein complex identification experiments~\cite{SupertiFuga-Complexes,Ho-Complexes} and iv) small-scale or module-scale experimental identification of protein complexes as reported in MIPS~\cite{MIPSdatabase} (Supp. Mat.).
We consider only the giant connected component produced by this data set, a network containing 1741 interactions amongst 741 proteins. Our data set choice reflects a desire to bias the data towards a thorough and accurate coverage of a limited region of the interactome as opposed to a wider, but likely shallower and more error prone, sampling. 
Note that when defining whether two proteins interact, a binary description is being imposed on what ideally would be characterized in terms of an affinity constant. Experimentally, effectively the aim has been for a cut-off that is high enough to exclude indiscriminate low-affinity interactions, such as those that occur between a general protein and proteasomal, ribosomal, or heat shock proteins, since {\em in principle} these are less informative interactions~\cite{SharedComponents, WeakLinks}. Hence, such interactions are largely absent from data sets.
With the vast majority of the proteins at the periphery of the map and well-defined modules connected through a handful of more central proteins, the yeast interactome polar map (Figure~\ref{MainFig}) presents what we term a `coordinated-functionality' architecture. Next, we discuss how this interactome architecture fits in with the biology of the cell.

An examination of the MIPS database functional (biological process) annotation of the proteins~\cite{MIPSdatabase} demonstrates that the topological modules make very good sense as biological functional modules.
On average 88.8\% of the proteins in a module share a similar function based on their MIPS classification~\cite{MIPSdatabase} (Supp. Mat.), confirming earlier studies connecting topology and function~\cite{FieldsModularity,GalitskiModularity,ChenModularity,VespignaniModularity,MirnyModularity,ZhouModularity,OuzounisModularity,GuenocheModularity,DunnModularity,MonaModularity}.
Part of the  mismatch between the topological modules and current protein functional annotations will likely vanish, once more complete and accurate interactome data sets become available.
However, it is the disparities not due to data set limitations that are truly interesting, for those are instances where the functional modules based on the interactome topology are not the ones we currently assign in functional classification schema.
In view of i) the good overall match found and ii) the foundational role of the interactome in the cell, we propose that the interactome topology represents a fundamental source for the division of proteins into functional groups. As such, its modularity analysis provides a rigorous alternative to the currently subjective functional annotation present in protein databases.
In accordance, we name the topological modules found so as to reflect their perceived biological roles (Figure~\ref{MainFig}).

The average degree of essential~\cite{Essentiality,Essentiality2} proteins in the data set is 5.7, while that of non-essential proteins is 3.9, a difference that is too large to be attributable to chance alone~\cite{JeongLethality} (Supp. Mat.).
This difference may indeed be biologically meaningful. Although physically knocking out a gene associated with a high or a low degree protein, say of degree 2 or 10 respectively, may be considered equivalent, from a mathematical network perspective it is not. In one case, it involves deleting 1 node {\em and} 2 links, in the other it involves deleting 1 node {\em and} 10 links. Note that such an explanation would not involve ascribing any out of the ordinary, higher-level role to hubs, the large degree proteins. Alternatively, the observed higher average degree of essential proteins could still stem from lingering systematic biases in the FYI data set.
 
While the translation of the interactome modular topology into biological functional modularity is straightforward, this is not the case for the interactome topological hierarchical organization. 
A fundamental question that the traffic analysis gives rise to is how the interactome topological hierarchical organization phenomenologically expresses itself. For instance, do more hierarchically central proteins in fact perform a higher-level, coordinating role in the cell? 
At present, these are open biological questions.

Comparing proteins of {\em equal} degree, we find no significant correlation between a protein's traffic and its essentiality~\cite{Essentiality,Essentiality2} (Supp. Mat.), something not too surprising, as knocking out a key protein that renders an essential functional module inoperant is plausibly more damaging than knocking out a protein that mediates two distinct processes that nonetheless can still function independently.
An intriguing, though at the moment still unsupported hypothesis is that, if a protein is disrupted, its traffic level correlates with the likelihood of causing non-lethal side effects in multiple areas of the biology of the cell. Speculating further, perhaps our representation of the interactome can provide clues as to where those side effects may arise --\, a matter of critical importance in drug development. 
A different possibility is that, for some of these module-connecting proteins, interacting with multiple modules is not a sign of a role in coordinating the functionality of the modules, but rather just a result of the protein being independently used in those modules. In opposition to a true functional `{\em connector}', we would call such a protein a `{\em bolt}' (alternatively, `widget'), in reference to how, analogously, a mechanical bolt can be used in multiple functional modules of a human engineered machine, while playing no role in coordinating their functionality.
Finally, note that {\em a priori} the observed 47 inter-module interactions are particularly susceptible to be false-positives, because a false-positive interaction between two random unrelated proteins is likely to result in an inter-module, high traffic interaction. However, significantly, 45 out of these 47 inter-module interactions belong to the set of interactions supported by small-scale targeted experiments and  arguably it is not very likely that a false-positive interaction of the type just described would go unnoticed in a targeted experiment and further make it into the peer-reviewed literature. Of the remaining 2 inter-module interactions, one is reported in the Ito {\em et al.}~\cite{ItoY2H}  and Uetz {\em et al.}~\cite{UetzY2H} high-throughput yeast two-hybrid data sets as well as in the MIPS data set of protein complexes identified via small or module scale experiments~\cite{MIPSdatabase}, while the other is reported in the Gavin {\em et al.}~\cite{SupertiFuga-Complexes} high-throughput protein complex identification  study and again in the Uetz {\em et al.}~\cite{UetzY2H} yeast two-hybrid data set. Out of the 45 supported by small-scale experiments, 4 are also reported in a high-throughput yeast two-hybrid study~\cite{UetzY2H,ItoY2H}, 3 in a high-throughput protein complex identification study~\cite{SupertiFuga-Complexes,Ho-Complexes} and 2 in the MIPS data set of protein complexes identified via small or module-scale experiments~\cite{MIPSdatabase}.
Whatever biological role central proteins turn out to play, we submit that they call for further experimental investigation, given their unique topological placement in the interactome.

\begin{figure}
\scalebox{0.70}{\includegraphics{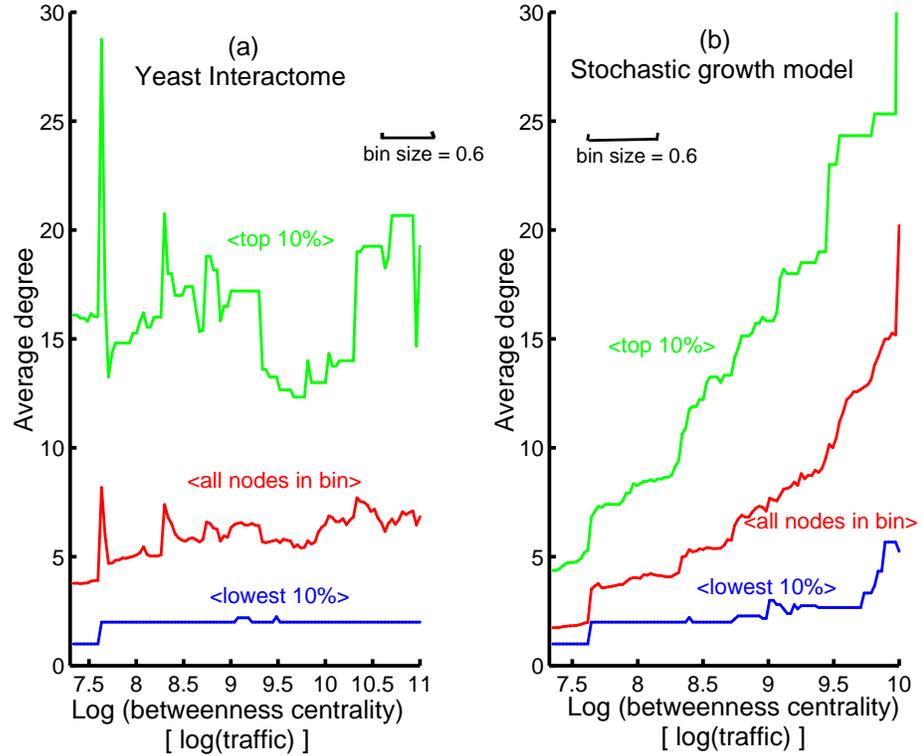}}  
\caption{\smaller{\em A non-hub-centric hierarchical organization.} {In the yeast interactome (a), the degree distribution does not change greatly as one moves from the periphery to the center of the Figure~\ref{MainFig} polar map (i.e, as one moves from a low to a high traffic region). In other words, hubs are not hierarchically central in the yeast interactome. In contrast, gene duplication and divergence interactome stochastic growth models~\cite{SoleModel,VespignaniModel,WagnerModel} produce hub-centric interactomes, where the average degree markedly increases with traffic and the hierarchical center of the interactome is therefore dominated by hubs. 
 {\bf (a)} Analysis for the yeast interactome giant connected component, based on the FYI data set (741 proteins, 1752 interactions)~\cite{Jackie}. Red curve - the average degree for the set of nodes whose $log\left(\mbox{traffic}\right)$ value falls within 0.3 of the $log\left(\mbox{traffic}\right)$ value indicated in the $x$-axis.
That is, a $log\left(\mbox{traffic}\right)$ bin of size 0.6 is continuously slid along the $log\left(\mbox{traffic}\right)$ axis and the average degree for all the nodes that fall within the bin is calculated. The last bin also includes all nodes with a $log\left(\mbox{traffic}\right)$ value larger than the range shown in the figure. The bins cover the entire data set, with every bin containing at least 26 nodes. The first bin contains 417 nodes. The last bin contains 37 nodes. 
Green curve - similar to the red curve, except this time the degree average is done only over the 10\% largest degree nodes in the bin. Blue curve - similar to green curve, but this time averaging over the 10\% lowest degree nodes in the bin. {\em The average degree in the highest traffic bin (rightmost data point in the red curve) is only 0.42 times the average degree of the 10\% largest degree nodes in the lowest traffic bin (leftmost data point in the green curve).}
{\bf (b)} Corresponding plots for the giant component of an interactome evolved under the gene duplication and divergence stochastic growth model of Pastor-Satorras {\em et al.}~\cite{SoleModel}. In this case, the giant component contains 759 nodes and 1542 interactions. Every bin contains at least 26 nodes. The first and last bins contain 432 and 40 nodes, respectively. {\em The average degree in the highest traffic bin is now 4.6 times larger than the average degree of the 10\% largest degree nodes in the lowest traffic bin.}
Similar results were achieved under multiple trials, model parameters and gene duplication growth models (Supp. Mat.).}}
\label{NoHubsFig}
\end{figure}

Having hierarchically classified the proteins with the traffic measure, we are now in a position to consider network degree distribution related questions without falling prey to previously noted statistical problems (Figure~\ref{HierarchyFig}).
Of particular interest is how degree changes as one moves hierarchically across the interactome~\cite{SuiHuang} (Figure~\ref{NoHubsFig}a).
Surprisingly, nodes of different degree are rather homogeneously hierarchically spread across the interactome: note the large spread between the green, red and blue curves {\em relative} to their small positive slopes; or, for a more quantifiable attribute, how the 10\% of nodes with the largest degree in the periphery of the interactome have a significantly larger degree than the average degree at the center of the interactome.
Thus, the interactome is not hierarchically stratified by degree. In particular, the interactome has a {\em non-hub-centric} hierarchical organization.
Further, note that should the inter-module protein interactions indeed represent a form of functional coordination, then this coordination is apparently occurring overwhelmingly in a pair-wise fashion: out of the 76 proteins that possess links to modules outside their own, only 4 connect their home module to more than one other module (TAF25 and SRP1 have links to 4 other modules, NUP1 and CLB2 to 2 other modules). The other 72 proteins connect their home module to a single other module (Supp. Mat.).
It is also noteworthy that, amongst these 76 connecting proteins, the higher degree ones in general belong clearly in their assigned home modules (specifically, for connecting proteins of degree 4 or higher, let us exclude the 2 interactions that each of these proteins must have by default to connect it to its home module and to one linked module; then, 95.3\% of the remaining interactions of connecting proteins are with the protein's respective home module. Supp. Mat.) 
This {\em pair-wise-coordination} is in sharp contrast with the picture of a hub protein connecting and mediating multiple modules~\cite{BarabasiBioReview,JeongLethality,BarabasiHierarchy}.

The non-hub-centric organization runs contrary to a number of network growth models that have been proposed to explain the topology of the interactome~\cite{BarabasiBioReview,SoleModel,VespignaniModel,WagnerModel,WigginsMechanisms}.
The models are based on evolution by stochastic gene duplication and divergence~\cite{AlbertsBook, OhnoBook}.
Amongst other reproduced statistics, the models are able to generate the power-law degree distribution observed in early interactome data sets~\cite{Doyle}. Since these models do not make appeal to evolutionary selection pressures, a major conclusion taken from their success was that natural selection is not required to reproduce the global structure of the interactome; instead, stochastic gene duplication and divergence suffices to give rise to that topology~\cite{SoleModel,WagnerModel}.
However, here we report that these gene duplication models lead to hierarchically hub-centric networks. In Figure 3b, we show data pertaining to an interactome built using the model of Pastor-Satorras {\em et al.}~\cite{SoleModel}. By comparison with the same plot for the yeast interactome, this time the network is clearly stratified by degree, with the larger degree nodes concentrated at the hierarchical center of the network. Now it is the average degree at the center of the network that is significantly larger than the average degree of the 10\% of nodes with largest degree at the periphery.
The models of V\'{a}zquez~\cite{VespignaniModel} (slightly different implementation of the gene duplication and divergence process), and of Wagner~\cite{WagnerModel} (emphasizing a continuous divergence in the form of gain and loss of interactions amongst existing proteins) produce similar hub-centric networks (Supp. Mat.).
In summary, there are at least three possible explanations for the non-hub-centric hierarchy we observe in yeast: i) it is a spurious effect associated with the limitations of existing data and the interactome is in fact hub-centric; ii) there is some crucial feature of the gene duplication and divergence mechanics that is not understood and/or is not captured by current models; or iii) the non-hub-centric hierarchy is in fact shaped by natural selection pressures {\em on the global} interactome structure.

Our study is based on the present-day knowledge of the yeast interactome, which is still rather deficient~\cite{OliverMain, OliverCorrelations,ItoReview, VonMeringComparison, BaderHogue, AloyRussell, GersteinStructural}.
We minimized false-positives by using the higher confidence yeast FYI data set as the source for our study.
The good correspondence between the topological modular breakdown of the interactome and the known functional annotation of proteins corroborates that false-positives are not an overriding problem in this data set.
Nonetheless, we repeated the interactome analysis using a data set of interactions reported in the MIPS database that are validated through small-scale screenings, the most reliable source of data~\cite{VonMeringComparison} (giant component: 392 proteins, 675 interactions. Supp. Mat.).
The correspondence between topological and functional modules (now on average 92\% of the proteins in a module shared a similar function based on their MIPS classification), as well as the reported non-hub-centric hierarchy were again supported by this small-scale data set (Supp. Mat.).
Regarding the limited coverage of our data sets (FYI $\!\approx\!12$\% coverage, small-scale $\!\approx\!7$\% coverage, assuming $\approx\! 6000$ proteins in yeast~\cite{ItoY2H}), it is of note that the doubling in size of the network, going from the  small-scale to the FYI data set, did not dilute the observed non-hub-centric topology nor the pair-wise inter-module connectivity pattern. 
In fact, the FYI data set produces an even slightly less hub-centric interactome than the small-scale data set (Supp. Mat.). Likewise, the pair-wise inter-module connectivity pattern is no less present in the FYI than in the small-scale data set (where out of the 47 proteins with inter-module links, 4 connect their home module to more than one other module. Supp Mat.). The observation that higher degree connecting proteins are in general strongly attached to their home modules is equally confirmed in both data sets (the earlier mentioned 95.3\% of home module interactions for the FYI higher degree connecting proteins, now becomes 95.7\% in the small-scale data set). 
Still, it is important to bear in mind the limitations of current data sets. For instance, regarding the typical pair-wise coordination, the possibility that this observation is only the result of a high-number of false-negatives for inter-module protein interactions cannot be ruled out. Note, for example, how in the map there are no interactions between the translation initiation module and the ribosomal subunit modules, even though such interactions must certainly exist.  
Ultimately, only the generation of more accurate and comprehensive interactome data sets can unequivocally confirm or disprove some of the results and hypotheses put forward in this article~\cite{OliverMain, OliverCorrelations}.

\begin{figure}
\scalebox{0.70}{\includegraphics{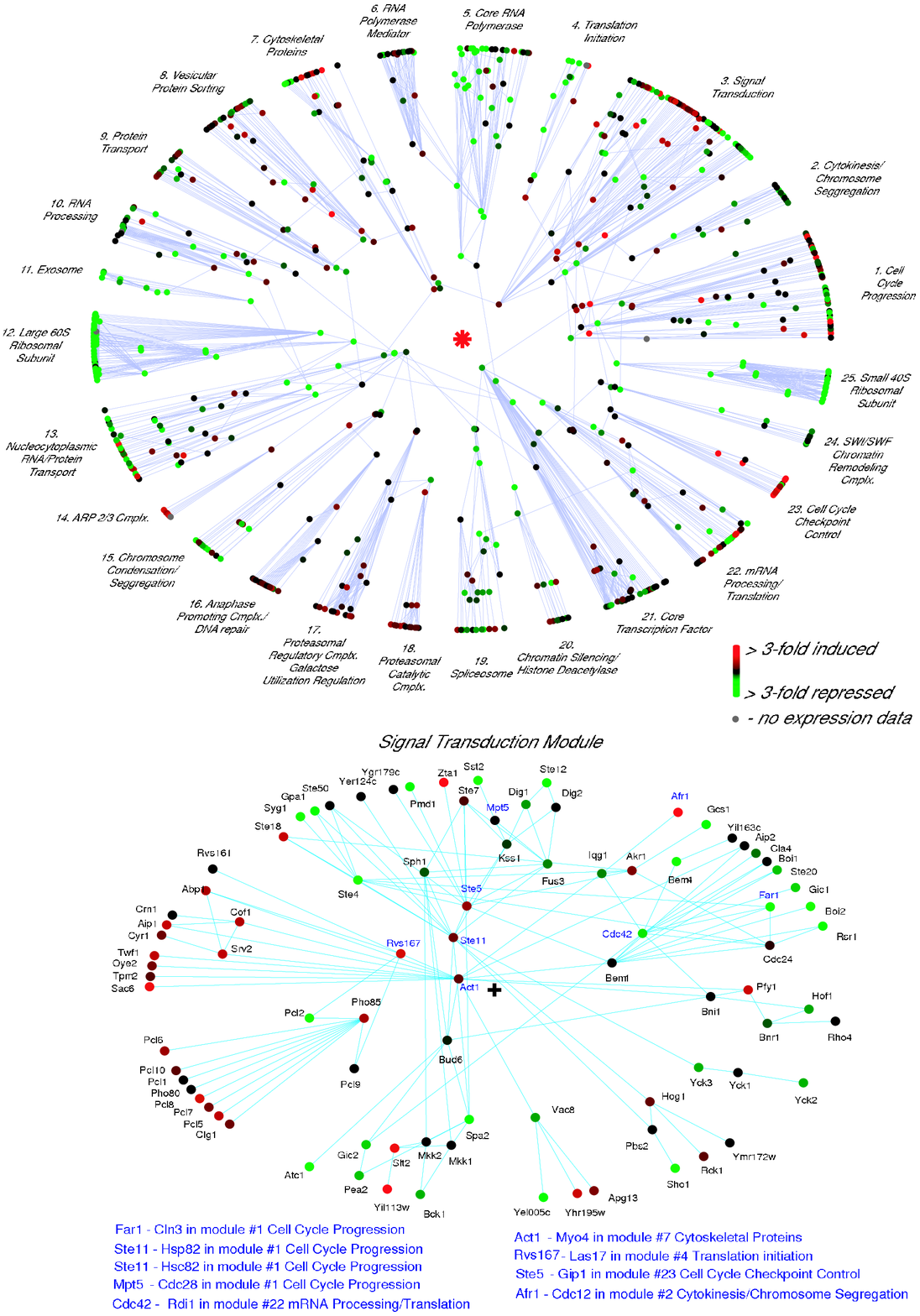}}
\vspace{-1.3 cm}
\caption{{\smaller\em Overlay of 20 min post heat shock mRNA expression data~\cite{Gasch} upon the interactome polar map.}
\smaller{{Maps are constructed as in Figure~\ref{MainFig} and Supp. Mat. Fig.~8 (proteins connected to other modules in the global map have names in blue). In addition, node color indicates mRNA expression level.
As previously reported~\cite{Gasch}, there is a sharp decline in expression of mRNA for ribosomal proteins. Other modules with a significant proportion of constituent proteins showing a decline are the Translation Initiation and Core RNA Polymerase modules. These observations are consistent with the repression of genes involved in RNA and protein synthesis upon heat shock~\cite{Gasch}. Modules with a significant proportion of constituent proteins showing an increase in expression are the small Cell Cycle Checkpoint Control module and parts of the Signal Transduction and Cell Cycle Progression modules. Overlay of mRNA expression data upon the single module maps of the latter two modules shows clear over expression in the protein chaperone submodule of Cell Cycle Progression (Supp. Mat. Fig.~9), and into the phosph-cyclin and cell growth/morphogenesis submodules of Signal Transduction (above). The increased expression of genes in these submodules supports the notion that cell growth is checked upon heat shock.}}}
\label{mRNAfig}
\end{figure}

So far our analysis has focused on the global interactome topology.
Now we would also like to highlight its potential as a framework for exploiting the wealth of interactome data. The interactome can form a valuable platform for crystalizing biological thought. 
We briefly introduce two relevant extensions to our work.
First, one may zoom into a module of interest in the interactome and locally repeat the analysis, producing a {\em single module polar map} (Figures in Supp. Mat.). Such a local map provides one with a starting point to discuss the biology of the process under study, interpret and design experiments, and generate new biological hypotheses.
Second, we note that the entire proteome is rarely, if ever, evenly expressed by the cell~\cite{AlbertsBook}.
Therefore, perhaps the interactome is best viewed as a potential network at the cell's disposal, with different parts of it being turned on and off to different degrees, as biologically required. Integrating mRNA expression data with the interactome polar map~\cite{mrnaBork} (Figure~\ref{mRNAfig}), permits a proper, unified analysis of this dynamical network.

The interactome represents an elementary abstraction of the multitude of complex biochemical interactions taking place in the rich physiological environment of the cell.
In the trade-off between simplicity and realism, arguably some facts may be beneficially incorporated in future interactome models: For instance, protein interactions vary within a continuum of binding affinities and post-translational modifications as well as allosteric interactions effectively change the possible binding partners of a protein, to name a few of the more prominent omissions at present~\cite{AlbertsBook}.
We end by noting that the organization of a network through a procedure akin to the one used in this paper may also be of relevance to the problem of network motif finding~\cite{AlonMotifs}, as certain motifs may turn out to occur sparsely overall and yet be statistically significant in specific regions of the network (for example, only in the high traffic central area, or only in a particular module).
\\

{\smaller Acknowledgments - A. Valente  thanks above all R. Fagerstrom, but also A. Sarkar S. Milstein, P.-O. Vidalain, M. Drezje, J. Han, S. Tee, K.-H. Lin and M. Boxem for comments and support, as well as the entire Vidal lab at Dana-Farber for its hospitality during part of the year. Finally, I am indebted to Howard Stone at Harvard University, under whose orientiation I initiated this work,  and to Phil Prorok, Gary Gao and Xiaoxia Lin for their unconditional support at a critical stage of this work. This work was supported by DFCI Sponsored Research.
}

\bibliography{scibib}

\bibliographystyle{Science}

\end{document}